\documentclass[pdflatex,sn-nature]{sn-jnl}

\usepackage{graphicx}%
\usepackage{multirow}%
\usepackage{amsmath,amssymb,amsfonts}%
\usepackage{amsthm}%
\usepackage{mathrsfs}%
\usepackage[title]{appendix}%
\usepackage{xcolor}%
\usepackage{textcomp}%
\usepackage{manyfoot}%
\usepackage{booktabs}%
\usepackage{algorithm}%
\usepackage{algorithmicx}%
\usepackage{algpseudocode}%
\usepackage{listings}%
\usepackage{caption}

\begin{document}

\title[Article Title]{Close-in planet induces flares on its host star}


\author*[1,2]{\fnm{Ekaterina} \sur{Ilin}}\email{ilin@astron.nl}

\author[1,3]{\fnm{Harish K.} \sur{Vedantham}}\email{vedantham@astron.nl}

\author[2,4]{\fnm{Katja} \sur{Poppenh\"ager}}\email{kpoppenhaeger@aip.de}

\author[1,3]{\fnm{Sanne} \sur{Bloot}}\email{bloot@astron.nl}

\author[1,5]{\fnm{Joseph R.} \sur{Callingham}}\email{callingham@astron.nl}

\author[6]{\fnm{Alexis} \sur{Brandeker}}\email{alexis@astro.su.se}

\author[7]{\fnm{Hritam} \sur{Chakraborty}}\email{hritam.chakraborty@unige.ch}

\affil*[1]{\orgdiv{Netherlands Institute for Radio Astronomy}, \orgname{ASTRON}, \orgaddress{\street{Oude Hoogeveensedijk 4}, \city{Dwingeloo}, \postcode{7991 PD}, \country{The Netherlands}}}
 
\affil[2]{\orgdiv{Leibniz-Institut für Astrophysik Potsdam}, \orgname{AIP}, \orgaddress{\street{An der Sternwarte 16}, \city{Potsdam}, \postcode{14482}, \country{Germany}}}

\affil[3]{\orgdiv{Kapteyn Astronomical Institute}, \orgname{University of Groningen}, \orgaddress{\street{P.O. Box 800}, \city{Groningen}, \postcode{9700 AV}, \country{The Netherlands}}}

\affil[4]{\orgdiv{Institut f\"ur Physik und Astronomie}, \orgname{Universit\"at Potsdam}, \orgaddress{\street{Karl-Liebknecht-Str. 24/25}, \city{Potsdam-Golm}, \postcode{14476}, \country{Germany}}}

\affil[5]{\orgdiv{Anton Pannenkoek Institute for Astronomy}, \orgname{University of Amsterdam}, \orgaddress{\street{Science Park 904}, \postcode{1098\,XH}, \city{Amsterdam}, \country{The Netherlands}}}

\affil[6]{\orgdiv{Department of Astronomy}, \orgname{Stockholm University}, \orgaddress{\street{AlbaNova University Center}, \city{Stockholm}, \postcode{10691}, \country{Sweden}}}

\affil[7]{\orgdiv{Geneva Observatory}, \orgname{University of Geneva}, \orgaddress{\street{Chemin Pegasi 51}, \city{Versoix}, \postcode{1290},  \country{Switzerland}}}

\newcommand{\hip}{HIP~67522\;}
\newcommand{\perday}{d$^{-1}$}
\newcommand{\emin}{$1.0\times10^{34}\,$erg}
\newcommand{\emax}{$7.4\times10^{35}\,$erg}

\abstract{In the past decade, hundreds of exoplanets have been discovered in extremely short orbits below 10 days. Unlike in the Solar System, planets in these systems orbit their host stars close enough to disturb the stellar magnetic field lines~\citep{saur2013magnetic}. The interaction can enhance the star's magnetic activity, such as its chromospheric~\citep{shkolnik2003evidence} and radio~\citep{pineda2023coherent} emission, or flaring~\citep{cohen2011dynamics}. So far, the search for magnetic star-planet interactions has remained inconclusive. Here, we report the first detection of planet-induced flares on HIP\,67522, a 17 million-year-old G dwarf star with two known close planets~\citep{rizzuto2020tess, barber2024tess}. Combining space-borne photometry from TESS and dedicated CHEOPS observations over a span of 5 years, we find that the 15 flares in HIP 67522 cluster near the innermost planet's transit phase, indicating persistent magnetic star-planet interaction in the system. The stability of interaction implies that the innermost planet is continuously self-inflicting a six time higher flare rate than it would experience without interaction. The subsequent flux of energetic radiation and particles bombarding \hip\,b may explain the planet's remarkably extended atmosphere, recently detected with the James Webb Space Telescope~\citep{thao2024featherweight}. HIP\,67522 is therefore an archetype to understand the impact of magnetic star-planet interaction on the atmospheres of nascent exoplanets.}




\maketitle

\hip is a young, 17 million year old star-planet system in the Upper Centaurus Lupus part of the Sco-Cen OB association~\citep{ dezeeuw1999hipparcos,rizzuto2020tess} about $125\,$pc away~\citep{gaiacollaboration2021gaiaa}. The star is a $1.2M_\odot$ dwarf~\citep{rizzuto2020tess} that hosts one gas giant planet in a $6.95\,$d orbit~\citep{rizzuto2020tess}, and another in a $14.33\,$d orbit~\citep{barber2024tess}.

\begin{figure}[h!]
\centering
\includegraphics[width=0.65\textwidth]{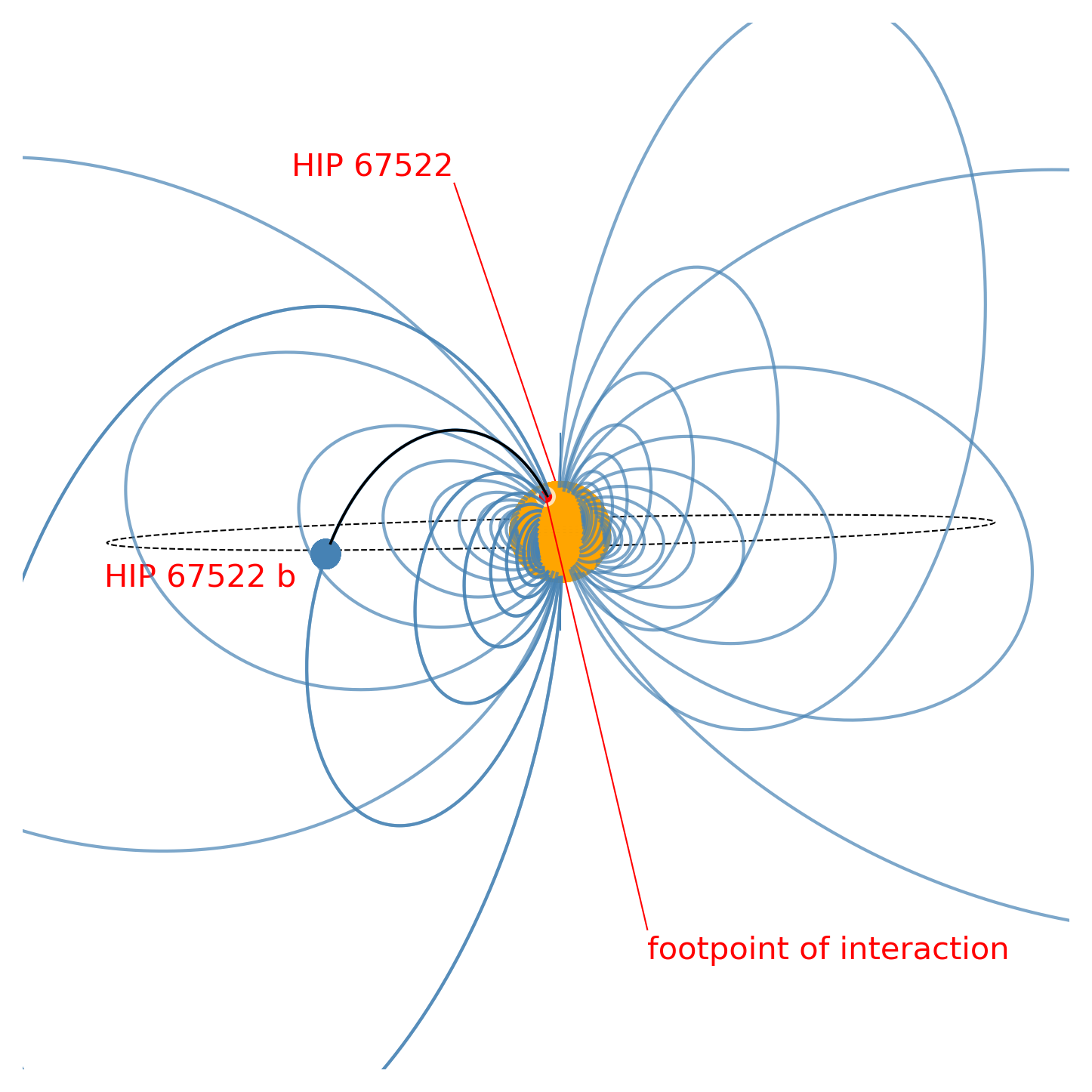}
\caption{Planet-induced flaring in the \hip system.  \hip b is shown as it perturbs the star's inclined magnetic field. The perturbation travels along the magnetic field line highlighted in white, toward the stellar surface, where it triggers flares at the footpoint, which is periodically visible to the observer. }\label{fig:toymodel}
\end{figure}

The innermost planet orbits less than $12$ stellar radii away from the star~\citep{rizzuto2020tess}, likely placing it in the sub-Alfv\'enic regime~\citep{saur2013magnetic, ilin2024planetary}. In this regime, perturbations of the stellar magnetic field induced by the planet can travel back to the star along the magnetic field lines that tether the two bodies~\citep{saur2013magnetic}~(Fig.~\ref{fig:toymodel}). The expected power of magnetic star-planet interaction of \hip with its innermost planet is about $10^{26}\,$erg/s -- among the highest powers expected from systems with known close-in companions~\citep{strugarek2022moves}. In other star-planet systems, magnetic interaction has been suggested to power chromospheric hot spots~\citep{shkolnik2003evidence}, polarized radio emission~\citep{vedantham2020coherent, pineda2023coherent}, and flares~\citep{fischer2019timevariable}, but none could be validated in follow-up observations, potentially due to changing magnetic field properties throughout stellar activity cycles~\cite{shkolnik2008nature}. Lacking reliable detections of planet-induced emission so far, the underlying mechanism of energy dissipation remains poorly understood. However, regardless of how the energy is deposited in the star's atmosphere, the fingerprint of magnetic star-planet interaction is the occurrence of emission that is modulated in phase with the orbit of the interacting planet. This periodic signature is unique as long there is no other comparable periodicity in the system, e.g., the rotation of the star.

\begin{figure}[h]
\centering
\includegraphics[width=0.65\textwidth]{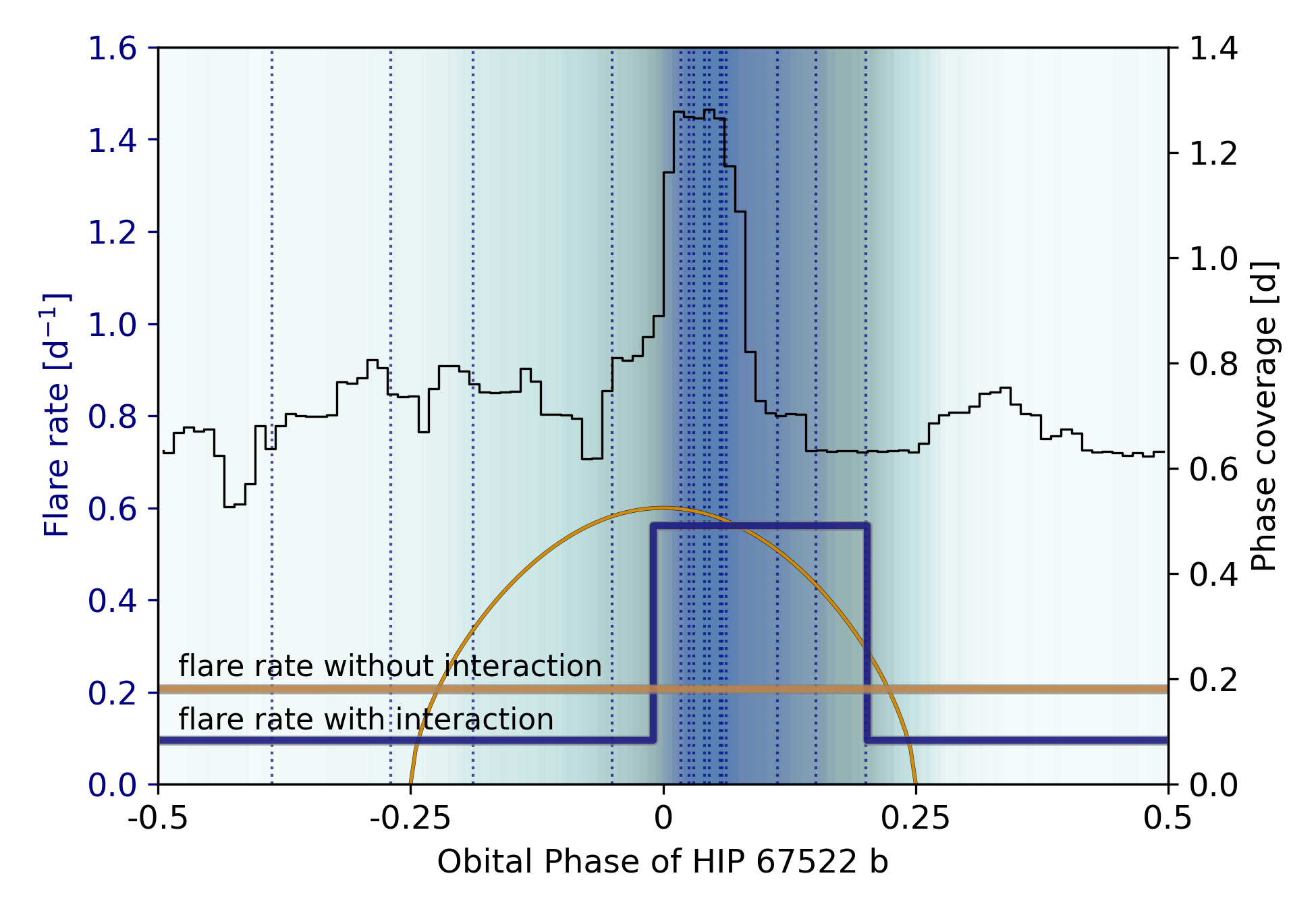}
\caption{Best-fit models with (dark blue) and without (orange) planet-induced flaring. The blue shade indicates at which orbital phases an elevated flare rate can be expected based on the observed flares (dotted lines) and the respective phase coverage (black line). The darker the shade, the more likely flaring is elevated at that orbital phase. The relative expected rate of planet-induced flares at the sub-planetary point (orange arc) at different orbital phases is consistent with the measured phase range of elevated flaring.}\label{fig:bestfit}
\end{figure}

\hip was previously observed by the Transiting Exoplanet Survey Satellite (TESS~\citep{ricker2015transiting}) in Sectors 11 (May 2019), 38 (May 2021), and 64 (April 2023) at a 2-min cadence for a total of $65.6\,$d~(Extended Data Fig.~\ref{ExtDat_fig:tess}). Additionally, we gathered dedicated observations of the \hip system with the CHaracterising ExoPlanets Telescope (CHEOPS~\citep{benz2021cheops}) for a total of $7.6\,$d between March 9 and June 22, 2024~(Extended Data Table~\ref{ExtDat_cheopsobslog}). We achieved excellent phase coverage: each hour of the $167\,$h orbit of \hip\,b was fully covered at least seven times~(Fig.~\ref{fig:bestfit} and Extended Data Table~\ref{ExtDat_obsphase}). We detected 12 flares in the TESS light curves~(Extended Data Fig.~\ref{ExtDat_fig:tessflares}), and 4 in the CHEOPS light curves (Extended Data Fig.~\ref{ExtDat_fig:cheops}, see also Methods). Of the CHEOPS flares, 3 were above the detection threshold for TESS. We use these for analysis, giving a total of 15 flares~(Extended Data Table~\ref{ExtDat_flares}), to ensure that the different detection thresholds for flares in CHEOPS and TESS do not bias our conclusion.

\begin{figure}[h]
\centering
\includegraphics[width=0.75\textwidth]{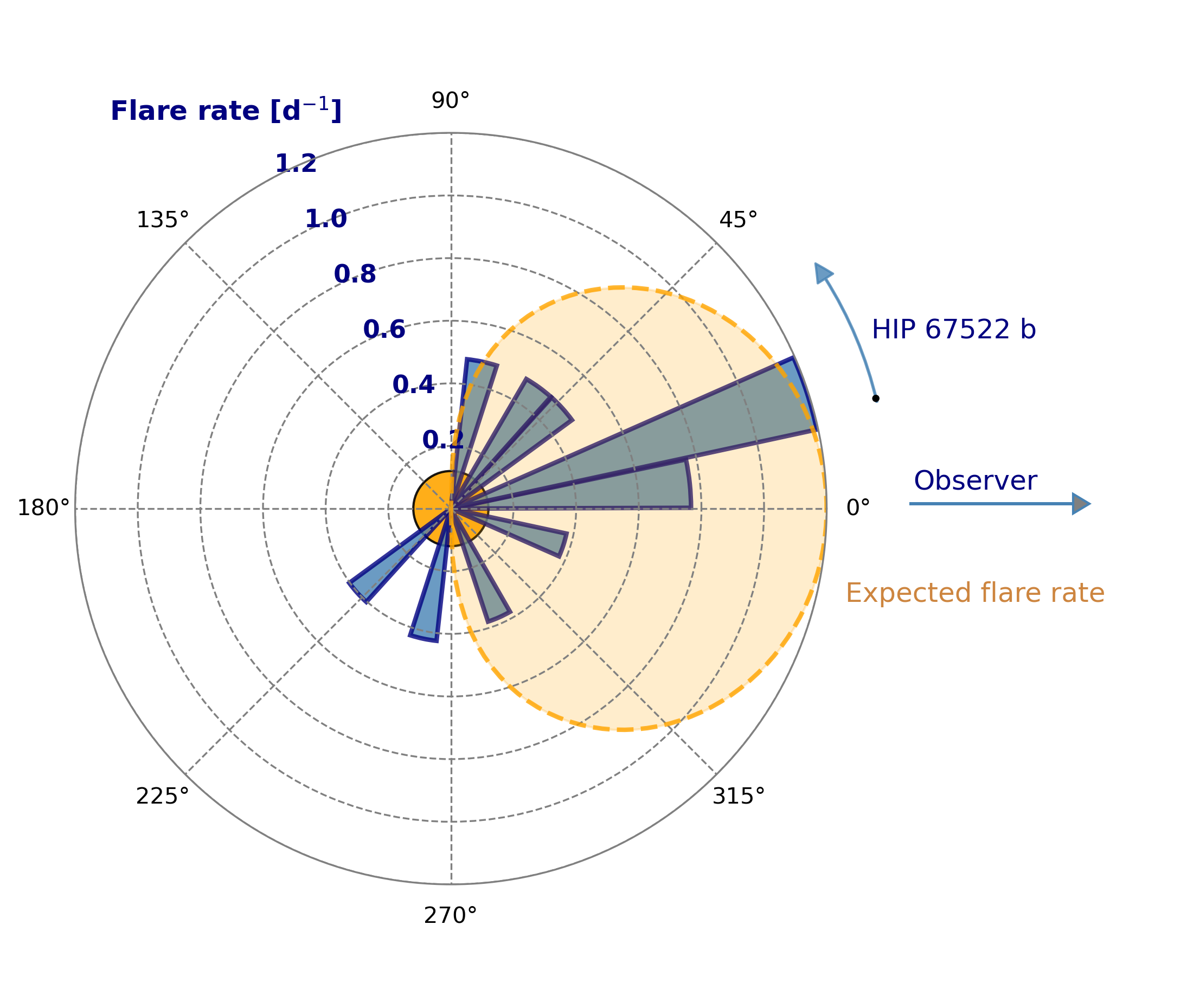}
\caption{Observed flare rates on \hip relative to the orbit of the innermost planet \hip\,b. \hip showed a significant increase in flare rate in the expected range, indicating that the majority of detected flares were triggered by the planet. The orange dashed region illustrates the expected rate of planet-induced flares from a footpoint of interaction at the sub-planetary longitude. Planet-induced flares can only be seen when the planet faces the observer ($\pm 90^\circ$), and most flares can be expected near the transit of \hip\,b ($0^\circ$). }\label{fig:polar}
\end{figure}

Fig.~\ref{fig:polar} shows the flare rate computed from the 15 flares as a function of the orbital phase of \hip\,b. The flare rate is highly elevated shortly after transit: we detect 11 flares in the orbital phase range 0--0.2 and only 4 flares throughout the rest of the orbit, amounting to an almost nine times higher flare rate in the 0--0.2 phase range. We confirmed the statistical significance of this elevated flare rate with a simple calculation. The 4 flares yield a base flare rate of $0.07$ flares per day, accounting for the observing time spent on the 0--0.2 range. Based on this flare rate, we anticipate an average of only $1.3$ flares in the $20\%$ range. Detecting 11 flares or more by chance at this base rate can be excluded with high confidence ($p<0.001$~\citep{gehrels1986confidence}). 
To ensure that our analysis is not biased by an a posteriori choice for the preferred 0--0.2 phase range, we implemented a Bayesian model selection that accepts the preferred range of elevated flaring as a free parameter. For this analysis, we calculated the total observing time per orbital phase bin, ensuring that the bin size is smaller than the clustering range, covering no more than $2\%$ of the orbit in each bin. We then compare two models that may explain the number of flares detected in each of these bins. The first model represents intrinsic stellar flaring without any star-planet interaction. In this model, flares have no dependence on orbital phase, such that the Poisson occurrence rate of flares in each time bin is the same. The second model allows a section of the orbit to show a different flare occurrence rate at the cost of three additional parameters in the second model, i.e., the second flare rate, the starting phase, and the width of the section with modulated flaring. Despite the added complexity, we find that the second model is strongly preferred over the first, with the ratio of the two marginal likelihoods, i.e., the Bayes Factor, of $11.7$, and a difference in the Akaike Information Criterion between the models of $7.5$. We also checked that these results were robust against variation in the bin sizes down to $0.5\%$ of the orbit (Methods, Extended Data Figure~\ref{ExtDat_fig:kaic}).

Our Bayesian analysis shows that the flare rate is elevated by a factor of $6$ for $20\%$ of the orbit of \hip\,b, beginning around transit. This orbital range is consistent with the clustering range that is expected considering the geometry of the system~(orange shadow in Fig.~\ref{fig:polar}): The planet orbits at low eccentricity~\citep{ rizzuto2020tess,barber2024tess}, and the stellar rotational and planetary orbital axes are aligned~\citep{heitzmann2021obliquity} (Methods). The visibility of flares at the footpoint of interaction for the observer is modulated by the optical self-shadowing of flares. Planet-induced flares can be best detected when the footpoint is closest to the center of the stellar disk, and worst near the limb. Additionally, flares cannot be detected when the footpoint is behind the stellar disk as seen from the Earth. If the large-scale magnetic field is well-described by a dipole field aligned with the rotation axis nearly in the plane of the sky~\citep{rizzuto2020tess,heitzmann2021obliquity}, the clustering would center exactly on transit, and the least flares would be observed when the planet is behind the star from the point of view of an observer on the Earth. However, as any cool dwarf, \hip also produces energetic flares intrinsically, so the flare rate need not be zero when the footpoint of interaction is not visible.

\begin{figure}[h!]
\centering
\includegraphics[width=0.75\textwidth]{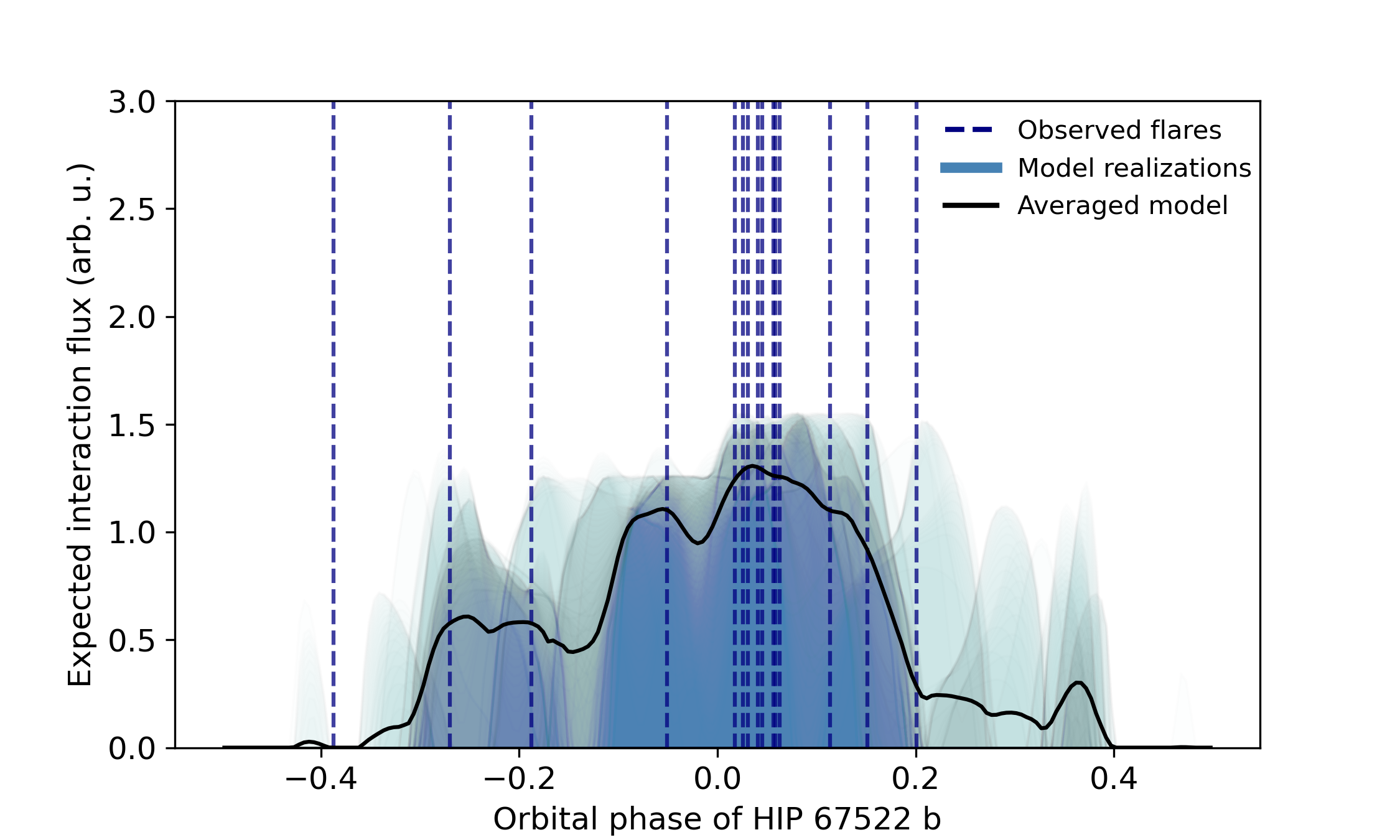}
\caption{Realizations of the geometric model under spin-orbit commensurability that peaks in the observed phase range. Some planet-induced flaring is expected outside the peak range, encompassing all but one flare in our sample.}\label{fig:toymodel_realizations}
\end{figure}

The best-fit center of the clustering region is at phase $0.08\pm0.04$, consistent with being centered on transit within $2\sigma$. The slight offset from transit may therefore be coincidental. Alternatively, it may be due to a distinctive feature of the \hip system. With a rotation period of approximately $1.42\,$d, \hip\,b orbits near a 1:5 commensurability between the rotation period of the star and the orbital period of \hip\,b. If this resonance was exact, the planet could align with the same stellar longitude at fixed orbital phases. In this scenario, a misaligned stellar magnetic field (Fig.~\ref{fig:toymodel}) would cause the planet to encounter periodic maxima and minima in magnetic field strength multiple times per orbit. This varying field induces harmonic modulation in the star-planet interaction strength, which depends on the position of the footpoint, and the stellar magnetic field strength encountered by the planet~\citep{saur2013magnetic, lanza2012starplanet}.  This harmonic modulation introduces two parameters into the model: the dipole field obliquity $\theta$ and the initial dipole azimuthal orientation $\phi_0$ relative to the observer. These parameters reproduce the peak power of interaction in the observed phase range for about $17\%$ of all possible configurations, spanning all obliquities above $20^\circ$~(Fig.~\ref{fig:toymodel_realizations}). If the system is not in resonance, this modulation would average out over time, leaving the clustering determined only by self-shadowing and visibility of the footpoint, similar to Fig.~\ref{fig:polar}. 

The detection of magnetic star-planet interaction in \hip is not a coincidental effect of testing a large sample of candidates. Among the few dozen flaring star-planet systems known to date, less than ten host a planet potentially close enough to be located within the sub-Alfv\'enic zone of the star's magnetosphere, and induce flares on its host star~\citep{ilin2024planetary}. Among those few, \hip has one of the highest predicted powers of magnetic interaction~\citep{ilin2024planetary}. 

The bolometric flux in the planet-induced flares on \hip is around $4.6\cdot 10^{29}\,$erg/s (Methods, Extended Data Fig.~\ref{ExtDat_fig:spiflux}). The available energy flux in the Alfv\'en-wing framework~\citep{saur2013magnetic} depends on the planetary magnetic field strength $B_p$ and stellar wind parameters. While stellar wind parameters can be measured in some cases~\citep{vidotto2021evolution}, no planetary magnetic fields have been reliably detected to date. Regardless, we find that the measured planet-induced flux cannot be supplied in the Alfv\'en-wing framework even with liberal assumptions on the planetary magnetic field strength and stellar wind parameters (Methods). We therefore suggest that the power in planet-induced flares in \hip is supplied by the energy reservoir stored in the star's coronal loops, with the star--planet interaction being a flare triggering mechanism. Such a mechanism would have to tap about $15\%$ percent of the $3 \cdot 10^{30}\,$erg/s coronal X-ray luminosity of \hip~\citep{maggio2024xuv} in order to match the measurement. We note, however, that the X-ray luminosity was derived from observations conducted shortly before the transit of \hip\,b, and may therefore represent a level of emission already elevated by (flaring) interaction.

Regardless of the underlying mechanism, the observed planet-induced emission has striking consequences. The orbital phases of the flares induced by \hip\,b imply that the interaction with the star backfires on the planet through self-inflicted space weather. A planet's space weather denotes the bombardment with high energy radiation and particles that erodes its atmosphere, and determines its ultimate mass and radius over cosmic time scales. Based on our results, \hip\,b is continually exposed to a roughly six times higher flare rate than it would be without interaction. Observations with the James Webb Space Telescope suggest that the bulk density of \hip\,b is extremely low: Despite its Jupiter-sized radius, its mass was constrained to less than $0.05M_{\rm Jup}$, with a best-fit mass of $15M_\oplus$~\citep{thao2024featherweight}. The low mass and strong irradiation by its young host star are considered the main reasons for the high inflation of \hip\,b's atmosphere~\citep{thao2024featherweight, maggio2024xuv}. The total amount of mass lost in the upcoming 100 million years will determine whether \hip\,b becomes a hot Neptune type planet, or will further erode and lose a significant fraction of its atmosphere to become a smaller, sub-Neptune object~\citep{thao2024featherweight}. Our results suggest that previous estimates, although already indicating considerable mass loss~\citep{thao2024featherweight, maggio2024xuv} from high energy radiation, may still be underestimating the total mass loss rate that can occur in \hip\,b under the exposure to the elevated flare rate. On the Sun, large flares are associated with coronal mass ejections (CMEs). On other stars, a CME is estimated to carry roughly the same the energy as the flare itself~\citep{moschou2019stellar}. If a CME is launched from the footpoint of flaring interaction near the sub-planetary point, it is much more likely to hit the planet than if the CME was triggered in random locations on the star. Under energy-limited escape, with a typical heating efficiency of the planet atmosphere of $10\%$ and CME angular widths typically observed from the Sun ($50^{\circ}-80^{\circ}$~\citep{jang2016comparison}), the impact of planet-induced CMEs would increase the mass loss rate derived from quiescent X-ray und UV flux~\citep{thao2024featherweight} by $50-130\%$~(Extended Data Fig.~\ref{ExtDat_fig:massloss}). Consequently, self-induced flaring may considerably shorten the lifetime of the planet's atmosphere from $1000$ to about $400-700$ million years.

Our results establish \hip as the archetype system for flaring star-planet interaction. \hip is the first star-planet system where the interaction persisted over a minimum of three years. The reliability of our detection opens the door to follow-up observations, and urges the characterization of \hip~c to allow for differential analysis of the two planets in the system. The well-known properties of \hip and \hip\,b allow us to target similar systems with a high probability of such interaction, such as V1298 Tau and TOI-837. With deeper understanding of the sub-Alfv\'enic conditions under which planets in planets like \hip\,b reside, magnetic star-planet interaction will not only probe exoplanetary space weather, but will also be instrumental for constraining their elusive magnetic fields~\citep{saur2013magnetic, lanza2012starplanet}. Planetary space weather and magnetic fields are crucial for understanding their atmospheric dynamics, assessing habitability, and probing interior compositions -- key to unraveling how planets throughout the Milky Way, including those in our Solar System, have formed and evolved.

\section*{Methods}\label{methods}

\subsection*{Photometric Data Reduction}

For the TESS photometry, we used the \texttt{PDC\_SAP} 2-min cadence light curves reduced by the Science Processing Operations Center (SPOC), available from the Mikulski Archive for Space Telescopes (MAST). 
For the CHEOPS photometry, we obtained 10\,s imagette exposures~(AO-4 ID4, PI: E.\,Ilin; AO-4 ID17, PI: H.\,Chakraborty), and used the open source photometric extraction package PIPE\footnote{\url{https://github.com/alphapsa/PIPE}} v1.1, designed to enable point-spread function (PSF) photometric extraction from the CHEOPS imagettes~\citep{brandeker2024pipe}. PSF photometry also reduces the roll modulation of CHEOPS data caused by the field of view rotating once per orbit in combination with background stars and an asymmetric PSF. We derived custom PSF models using data from the observations themselves, as described by the PIPE manual\footnote{\url{https://github.com/alphapsa/PIPE/blob/4d592ac796e57a7accfb78bbea96577b90eceb71/docs/pipe/PIPE_manual.pdf}}.  We removed flagged data points from all TESS and CHEOPS light curves prior to searching them for flares.

\subsection*{Flare Detection}
Flare detection follows three steps: removal of astrophysical variability, automatic identification of flare candidates, and confirmation of bona fide flares through visual inspection and ancillary data.

We removed the variability from the TESS light curves using the approach detailed in \citep{ilin2024planetary}. The CHEOPS data could not be treated in the same way because of the much shorter observing baselines and frequent gaps in the observations. We modeled each $13-23\,$h CHEOPS visit individually with a combination of the \texttt{batman}~\citep{kreidberg2015batman} transit model for \hip\,b with quadratic limb darkening using the planetary orbital and limb darkening parameters from~\cite{barber2024tess}, and a 5th degree polynomial to approximate the star's rotational variability. Three large flares and all data points with a CHEOPS quality flag $>0$ were masked prior to fitting the model using a least-square fit (\texttt{scipy.curve\_fit},~\cite{mckinney2010data}). We then subtracted the model from the data, and masked all outliers above 4 standard deviations in the residuals. We then repeated the model fit with the updated mask to obtain the final model. The model was then subtracted again from the data, and we applied a Savitzy-Golay~\citep{savitzky1964smoothing} filter to smooth the remaining non-flaring variability. For the filter, we chose a 3rd order polynomial, and a window that was $20\%$ of the total length of each visit with CHEOPS. As a final correction, we subtracted the remaining roll dependent systematics that were not captured by PIPE using the median of the 100 closest data points in roll angle as the truth for any data point in the light curve. 

For the initial flare identification from the resulting flattened light curves, we followed the approach in \citep{ilin2024planetary}, where three consecutive data points three standard deviations above the noise level define a candidate event. The noise levels of the CHEOPS and TESS light curves are similar, $810\pm 90\,$ppm and $780\pm70\,$ppm at $10\,$s and $120\,$s cadence, respectively, defined as the standard deviation of the de-trended light curves.

The resulting candidates were manually inspected, for a total of four bona fide flares in the CHEOPS light curves, and twelve in the TESS light curves. We rejected several false positive candidates in CHEOPS, because their occurrence was invariably connected to a steep rise in background flux, or repeated roll angle modulation. In TESS, we found no false positives. We validated that the 16 final flares were indeed associated with the star by verifying the absence of moving objects in the field or variability associated with the entire image in the pixel level data.

\subsection*{Flare Energy}

We characterized each flare by fitting an empirical, analytic, continuous flare template~\citep{mendoza2022llamaradas} to its light curve. The baseline flux was established using the same technique used to de-trend the CHEOPS light curves. We first estimated the best parameters with a least-square fit (\texttt{scipy.curve\_fit},~\cite{mckinney2010data}), then used the results to efficiently sample the posterior distribution with the MCMC method (\texttt{emcee},~\cite{foreman-mackey2013emcee}), assuming flat priors for all flare parameters, i.e. the flare amplitude, full width at half maximum, and peak time of the flare. 

We calculated the bolometric energy of the flares following the prescription in~\cite{shibayama2013superflares}, which converts the flux emitted by the star and the flare in the TESS and CHEOPS passbands to the bolometric emission from the flare. We assumed the temperatures of both the star and the flare as known.
We propagated the posterior distribution of the flare parameters to derive the bolometric flare energy using Gaussian distributions for the stellar radius $R_* = 1.39\pm0.06$~\citep{rizzuto2020tess} and effective temperature $T_{\rm eff} = 5650\pm75$~\citep{rizzuto2020tess}. We assumed that the flares were black body emitters with a temperature of $10^4\,$K. We note that, on the one hand, the flare temperature can vary greatly~\citep{howard2020evryflarea,rabellosoares2022blackbody}. On the other hand, the flare energy is not very sensitive to the assumed flare temperature when using the prescription in~\cite{shibayama2013superflares}.

\subsection*{Characterizing the Clustering of Flares in Orbital Phase}

To ascertain that the clustering of flares was a clustering event, and not a confluence of random occurrences of intrinsic flares, we cast the two possible scenarios as Bayesian models, one with and one without flare clustering. The first model $M_{\text{unmod}}$ assumes the absence of any modulation, i.e., that no clustering is present, so that the data can be reproduced by a single Poisson flare rate $\lambda_0$. The rate of flares $x_i$ in a phase bin $i$ is then determined by the total observing time $t_i$ in that phase bin:

\begin{equation}
    x_i = \lambda_0 t_i
\end{equation}

The individual observed numbers of flares $n_i$ are independent of each other, so that the likelihood of a single flare rate $\lambda_0$ representing the full set of observations $D$ split into $k$ bins as $D=[n_1, .., n_k]$ reads:

\begin{equation}
    P(D | \lambda_0) = \Pi_{i=1}^k \dfrac{\text{e}^{-\lambda_0 t_i} (\lambda_0 t_i)^{n_i}}{ n_i!}
\label{eq:likelihood0}
\end{equation}
We have no prior information about $\lambda_0$, which we can represent as Jeffrey's prior:

\begin{equation}
    P(\lambda_0) =  \frac{1}{2\sqrt{2\lambda_0}}
\end{equation}
Jeffrey's prior is improper, i.e., the integral over all $\lambda_0\geq0$ is not finite, so we limit $0<\lambda_0<2\,$\perday, where the upper limit is above the highest flare rate found for stars with similar spectral type a fast rotation across the TESS sky in \citep{tu2020superflares} above the minimum detected flare energy in our flare catalog~(Extended Data Figure~\ref{ExtDat_fig:tu}). This upper limit is  conservative, and likely an overestimate because the sample in \cite{tu2020superflares} also includes extremely active BY Dra type variables. The $\sqrt{2}$ in the denominator is the normalization factor $N = \int_0^2 \lambda_0^{-1/2} \rm{d}\lambda_0$. Following Bayes' theorem, the probability of $\lambda_0$ given the data is

\begin{equation}
    P(\lambda_0 | D) = \dfrac{P(D | \lambda_0)  P(\lambda_0)}{P(D)},
\end{equation}
with the marginal likelihood

\begin{equation}
    P(D) = P(M_{\text{unmod}}) = \int_0^\infty \mathrm{d}\lambda_0 P(D | \lambda_0)  P(\lambda_0).
    \label{eq:marginal0}
\end{equation}
The second model, $M_{\text{mod}}$, assumes that the presence of a modulation, such that there is an orbital phase range $(\phi_0,\phi_0+\Delta \phi)$ during which the flare rate is $\lambda_1 \neq \lambda_0$, such that

\begin{equation}
    x_i =  \begin{cases}
\lambda_1 t_i & \phi_0<\phi_i<\phi_0+\Delta\phi,\text{ and}\\
\lambda_0 t_i &\text{else}.
\end{cases}
\end{equation}
We note that the phase bins $i$ must be small compared to the extents of the two phase ranges in order for each bin to be uniquely attributed to either range. The corresponding likelihood for the full set of parameters, $\Theta = [\lambda_0,
\lambda_1,\phi_0,\Delta \phi]$, 

\begin{equation}
    P(D | \Theta) = {\Pi_i^k} \begin{cases}  \dfrac{\text{e}^{-\lambda_1 t_i } (\lambda_1 t_i)^{n_i}}{ n_i!} & \phi_0<\phi_i<\phi_0+\Delta\phi\\
    \dfrac{\text{e}^{-\lambda_0 t_i } (\lambda_0 t_i)^{n_i}}{ n_i!} &\text{else} \end{cases},
\label{eq:likelihood1}
\end{equation}
and uninformative prior, 

\begin{equation}
    P(\Theta) =   2 \frac{1}{2\sqrt{2\lambda_0}}  \frac{1}{2\sqrt{2\lambda_1}},
\end{equation}
assume that $\lambda_0$ and $\lambda_1$ are independent, and take values between $0$ and $2$. We did not make assumptions about where the clustering begins ($\phi_0$) or how wide the clustering phase range is ($\Delta \phi$). Therefore, we let $\phi_0$ run from 0 to 1 with equal probability. We restrict $\Delta 
\phi$ from 0 to 0.5 so that we avoid double-counting the identical solution pairs generated by the substitutions: $\phi_0\rightarrow \phi_0+\Delta\phi$, $\Delta\phi\rightarrow1-\Delta\phi$, $\lambda_0\rightarrow \lambda_1$ and $\lambda_1\rightarrow\lambda_0$ (Note that we do not impose $\lambda_1>\lambda_0$). Normalization of the prior probability of $\Delta\phi$ contributes the factor of $2$ to the prior. The marginal likelihood for $M_{\text{mod}}$ is:

\begin{eqnarray}
     P(D) &=& P(M_{\text{mod}})  \nonumber\\ 
     &=&\int_{0}^{\infty} \mathrm{d}\lambda_0\int_0^\infty \mathrm{d}\lambda_1\int_0^1\mathrm{d}\phi_0\int_0^1\mathrm{d}\Delta\phi\, P(D| \Theta)  P(\Theta)
 \label{eq:marginal1}
\end{eqnarray}
We sampled the posterior distribution with \texttt{emcee}~\citep{foreman-mackey2013emcee} to obtain uncertainties on the best-fit parameters for the model~(see following Section). To decide which model better represents the data, we calculated the Bayes factor, that is the ratio of the two marginal likelihoods,

\begin{equation}
    K = \dfrac{P(M_{\text{mod}})}{P(M_{\text{unmod}})}.
\end{equation}
The Bayes factor penalizes the increase in the number of parameters in $M_{\text{mod}}$, as the integral must be performed over a much larger parameter space in Eq.~\ref{eq:marginal1} compared to Eq.~\ref{eq:marginal0}. The two equations can be solved numerically. We chose a sufficiently high grid size of 400, 400, 100, and 50 for $\lambda_0$, $\lambda_1$, $\phi_0$ and $\Delta \phi$. We further chose bin widths between $0.5\%$ and $2\%$ of the orbit, such that possible small widths of the clustering phase range $\Delta \phi$ could be sufficiently well sampled, and no phase information was lost due to coarse binning. The resulting Bayes factors varied from bin size to bin size, with a mean and standard deviation of $11.7\pm 1.9$. The average Bayes factor increases slightly toward smaller bin sizes, but the trend flattens out above 75 bins at $K=11.8\pm 1.6$.



Besides the Bayes factor, we also computed the Akaike Information Criterion (AIC) from the best-fit results obtained from sampling the posterior with \texttt{emcee}, where each best-fit parameters are defined as the $50$th percentile of the posterior distribution. The AIC is an approximation to the Bayes factor that compares the likelihoods of the best-fit solutions instead, while penalizing the number of parameters in the model. We tested bin widths between $0.5\%$ and $2\%$ of the orbit, for an average and standard deviation in the difference of AIC values between the two models of $7.5\pm1.4$ in favor of the modulated model, and $7.5\pm1.3$ above 75 bins.

\subsection*{Energetics}



In the unmodulated flaring model, the best-fit flare rate is $\lambda_0 = 0.21^{+0.06}_{-0.05}\,$\perday above the minimum detected flare energy of \emin, consistent with the simple division of 15 flares by the total observing time of $73.2\,$d. For the modulated flaring model, the flare rate outside the clustering range is lower, $\lambda_0 = 0.09^{+0.06}_{-0.04}\,$\perday. The planet-induced flaring rate is roughly six times higher than the base rate, at $\lambda_1 = 0.59^{+0.38}_{-0.22}\,$\perday. The flare rate is elevated starting at around transit, i.e., $\phi_0=0.00^{+0.02}_{-0.08}$, and lasts for about $20\%$ of the orbit, i.e., $\Delta\phi=0.20^{+0.11}_{-0.12}$.

We calculated the flux emitted in flares through planet-induced flaring by extrapolating the flare frequency distribution to lower and higher flare energies. We used the difference in flare rate $\lambda_1-\lambda_0$ and the power law slope $\alpha \approx 1.6$ derived from the flare frequency distribution of the full sample of 15 flares following the method in \cite{ilin2021flares}. Since the power law slope is $\alpha<2$, the total flux is dominated by high energy flares. For a conservative estimate, we set the maximum energy of planet-induced flares to the highest flare energy detected in the elevated flaring range, that is \emax. Flares at and above this energy occur on a monthly basis. With the chosen maximum energy, the total flux in planet-induced flares is $4.6\times 10^{29}\,$erg/s. Even if the three highest energy events out of the 11 flares in the elevated flaring region were not planet-induced but stemmed from the intrinsic $\lambda_0$ contribution, the planet-induced flux would still be $> 10^{29}\,$erg/s

The planet-induced flare flux is considerably higher than the interaction flux expected in the Alfv\'en wing model~\citep{saur2013magnetic}, even under very generous assumptions. Using the same procedure as in \cite{ilin2024planetary}, we calculate the average magnetic field strength of \hip from its X-ray luminosity from \cite{maggio2024xuv}, and alternatively, from its Rossby number, following \cite{reiners2022magnetism}, which yield $3.1$ and $2.1\,$kG, respectively. For planetary magnetic fields $B_p=0.01-100\,$G, stellar wind densities at the coronal base of $\rho = 0.1-100\rho_\odot$ ($\rho_\odot = 10^{10}\,$cm$^{-3}$), and stellar average magnetic field strengths $B=1600-3500\,$G, the star-planet interaction flux yields $10^{24}-10^{28}\,$erg/s, where the strongest flux is obtained with the highest values in the given ranges. Therefore, it is unlikely that the Alfv\'en wing interaction alone powers the elevated flaring in \hip\hspace{-0.1cm}.

\subsection*{Geometric Model of Flaring Star-Planet Interaction}

A clustering of flares in \hip in the observed phase range can be reproduced under minimal assumptions of the system's architecture and magnetic field. For this simple model, we assume a dipole magnetic field with some inclination $\alpha$ relative to the rotational inclination axis $i$, which we set to $i=90^\circ$, and zero inclination of the planetary orbit, consistent with observations~\citep{heitzmann2021obliquity}. We then assume that the large scale field of \hip is purely dipolar, and rotates with $P_{\text{rot}}=P_{\text{orb}}/5$. Any field line that intersects the planetary position and maps onto the surface of the star comes around every synodic period $P_{\text{syn}}$, i.e. the beat period between orbit and rotation, when the same stellar longitude faces the planet. In the observer frame, we account for the visibility of the footpoints on both hemispheres and the self-shadowing of the likely optically thick flare emission regions as they move in and out of view on the stellar surface. Of the two footpoints, the footpoint with the shortest path along the field line dominates, which is always the footpoint on the hemisphere facing the planet. The flux $F_{\rm SPI}$ we can expect to measure from the dominant footpoint is proportional to the magnetic field energy $\propto B^\beta$ with $\beta>0$~\citep{saur2013magnetic,lanza2012starplanet} at the position of the planet multiplied by the self-shadowing effect, $\cos\theta\cos\phi$, where $\theta$ and $\phi$ are the latitude and longitude of the footpoint of interaction. To convert to a value relative to the flare rate, $F_{\rm SPI}$ must be taken to the exponent $\alpha-1\approx 0.6$. Finally, we assume that \hip\,b is constantly orbiting within the sub-Alfv\'enic zone, so that interaction is possible anywhere along the orbit.

The self-shadowing of flares in the optical dominates the visibility of star-planet interaction, leading to the arc-shaped expected flare rate shown in Fig.~\ref{fig:polar} and~\ref{fig:bestfit}. If, in addition, the orbit of \hip\,b is an integer multiple of the star's rotation period, as is suggested for \hip\hspace{-0.1cm}, $P_{\rm syn}$ modulates the visibility of planet-induced flares, such that the observed shift away from transit can be reproduced. 

\backmatter


\section*{Declarations}

\textbf{Funding:} E.I. and H.K.V. acknowledge funding from the European Research Council under the European Union's Horizon Europe programme (grant number 101042416 STORMCHASER). K.P. acknowledges funding from the German \textit{Leibniz-Gemeinschaft} under project number P67/2018. S.B. acknowledges funding from the Dutch research council (NWO) under the talent programme (Vidi grant VI.Vidi.203.093). A.Br.\ was supported by the SNSA. H.C. acknowledge the support of the Swiss National Science Foundation under grant number PCEFP2\_194576
\newline
\textbf{Conflict of interest/Competing interests:} The authors declare no competing interests or conflict of interest.
\newline
\textbf{Data availability:} This paper includes data collected by the TESS mission, which are publicly available from the Mikulski Archive for Space Telescopes (MAST). Funding for the TESS mission is provided by the NASA's Science Mission Directorate. CHEOPS data analyzed in this article will be made available in the CHEOPS mission archive (\url{https://cheops.unige.ch/archive_browser/}). 
CHEOPS is an ESA mission in partnership with Switzerland with important contributions to the payload and the ground segment from Austria, Belgium, France, Germany, Hungary, Italy, Portugal, Spain, Sweden, and the United Kingdom. The CHEOPS Consortium would like to gratefully acknowledge the support received by all the agencies, offices, universities, and industries involved. Their flexibility and willingness to explore new approaches were essential to the success of this mission. 
\newline
\textbf{Code availability:} All code necessary to reproduce the results in this manuscpript is available on GitHub \url{https://github.com/ekaterinailin/hip67522-spi/tree/flaring-spi}.
\newline
\textbf{Author contributions:} E.I. and K.P. initiated the star-planet interaction search project.  E.I. led the processing of TESS and CHEOPS data with input from A.B., led the energetics and geometric calculations with input from H.K.V. and K.P.. E.I and H.K.V. performed the clustering analysis with input from J.R.C. and S.B.. S.B. provided input on the stellar wind estimate. H.C. aided with the data collection and sharing across projects. All authors commented on the manuscript. Correspondence and requests or materials should be addressed to E.I. (ilin@astron.nl).

Reprints and permission information is available at \url{www.nature.com/reprints}.

\noindent
%
%
%
%
%

\begin{appendices}

\section*{Extended Data}

\captionsetup[table]{name=Extended Data Table}
\captionsetup[figure]{name=Extended Data Figure}

\begin{table}[h]
\caption{CHEOPS observing log.}\label{ExtDat_cheopsobslog}%
\begin{tabular}{@{}llll@{}}
\hline
OBSID & File Key & Start Date [UTC] & Obs. baseline [h] \\
\hline
2365179 & CH\_PR240004\_TG000101\_V0300 & 2024-03-09 05:09 & 13.18 \\
2367052 & CH\_PR240004\_TG000102\_V0300 & 2024-03-16 05:56 & 13.18 \\
2370611 & CH\_PR240004\_TG000103\_V0300 & 2024-03-23 03:34 & 13.18 \\
2377976 & CH\_PR240004\_TG000104\_V0300 & 2024-03-30 02:46 & 13.18 \\
2406487 & CH\_PR240004\_TG000105\_V0300 & 2024-05-10 20:52 & 13.8 \\
2413239 & CH\_PR240004\_TG000106\_V0300 & 2024-05-17 19:31 & 13.19 \\
2421684 & CH\_PR240004\_TG000107\_V0300 & 2024-05-24 17:53 & 15.04 \\
2432394 & CH\_PR240004\_TG000108\_V0300 & 2024-05-31 18:07 & 13.18 \\
2444115 & CH\_PR240004\_TG000109\_V0300 & 2024-06-07 16:58 & 13.18 \\
2446774 & CH\_PR240004\_TG000110\_V0300 & 2024-06-14 17:24 & 14.17 \\
2455494 & CH\_PR240004\_TG000111\_V0300 & 2024-06-21 14:03 & 13.18 \\
2382459 & CH\_PR240017\_TG000101\_V0300 & 2024-04-05 15:37 & 19.98 \\
2382460 & CH\_PR240017\_TG000102\_V0300 & 2024-04-12 14:38 & 19.1 \\
2390044 & CH\_PR240017\_TG000103\_V0300 & 2024-04-19 13:41 & 22.77 \\
2402938 & CH\_PR240017\_TG000104\_V0300 & 2024-05-03 12:11 & 19.29 \\
2394579 & CH\_PR240017\_TG000501\_V0300 & 2024-04-28 17:25 & 15.53 \\
2383621 & CH\_PR240017\_TG000601\_V0300 & 2024-04-11 10:48 & 15.53 \\
2402943 & CH\_PR240017\_TG000701\_V0300 & 2024-04-30 12:18 & 15.53 \\
2366465 & CH\_PR240017\_TG000801\_V0300 & 2024-03-11 08:35 & 17.82 \\
2372811 & CH\_PR240017\_TG000901\_V0300 & 2024-03-21 03:03 & 16.7 \\
2435119 & CH\_PR240017\_TG001001\_V0300 & 2024-05-26 15:30 & 16.65 \\
 &  &  & $\Sigma_{\rm baseline}=$ 327.4 \\
 &  &  & $\Sigma_{\rm observed}=$ 183.0 \\
\botrule
\end{tabular}
\end{table}

\begin{table}[h]
\caption{Orbital phase coverage.}\label{ExtDat_obsphase}%

\begin{tabular}{@{}llll@{}}
\toprule
Orb. phase & Exposure time [d] & Orb. phase & Exposure time [d] \\
\midrule
(0.0, 0.01] & 0.997338 & (0.5, 0.51] & 0.493056 \\
(0.01, 0.02] & 1.239352 & (0.51, 0.52] & 0.486111 \\
(0.02, 0.03] & 1.260880 & (0.52, 0.53] & 0.529514 \\
(0.03, 0.04] & 1.251157 & (0.53, 0.54] & 0.657870 \\
(0.04, 0.05] & 1.251273 & (0.54, 0.55] & 0.670602 \\
(0.05, 0.06] & 1.274074 & (0.55, 0.56] & 0.685532 \\
(0.06, 0.07] & 1.204398 & (0.56, 0.57] & 0.729514 \\
(0.07, 0.08] & 1.158333 & (0.57, 0.58] & 0.732870 \\
(0.08, 0.09] & 0.924421 & (0.58, 0.59] & 0.732755 \\
(0.09, 0.1] & 0.664468 & (0.59, 0.6] & 0.733681 \\
(0.1, 0.11] & 0.626389 & (0.6, 0.61] & 0.739699 \\
(0.11, 0.12] & 0.626389 & (0.61, 0.62] & 0.738657 \\
(0.12, 0.13] & 0.625000 & (0.62, 0.63] & 0.759722 \\
(0.13, 0.14] & 0.627778 & (0.63, 0.64] & 0.759722 \\
(0.14, 0.15] & 0.626389 & (0.64, 0.65] & 0.765278 \\
(0.15, 0.16] & 0.627778 & (0.65, 0.66] & 0.763889 \\
(0.16, 0.17] & 0.626389 & (0.66, 0.67] & 0.708333 \\
(0.17, 0.18] & 0.626389 & (0.67, 0.68] & 0.694444 \\
(0.18, 0.19] & 0.625000 & (0.68, 0.69] & 0.695833 \\
(0.19, 0.2] & 0.626389 & (0.69, 0.7] & 0.665278 \\
(0.2, 0.21] & 0.625000 & (0.7, 0.71] & 0.658333 \\
(0.21, 0.22] & 0.625000 & (0.71, 0.72] & 0.722222 \\
(0.22, 0.23] & 0.626389 & (0.72, 0.73] & 0.738657 \\
(0.23, 0.24] & 0.626389 & (0.73, 0.74] & 0.734028 \\
(0.24, 0.25] & 0.626389 & (0.74, 0.75] & 0.721412 \\
(0.25, 0.26] & 0.626389 & (0.75, 0.76] & 0.721644 \\
(0.26, 0.27] & 0.655324 & (0.76, 0.77] & 0.732986 \\
(0.27, 0.28] & 0.677315 & (0.77, 0.78] & 0.736806 \\
(0.28, 0.29] & 0.695602 & (0.78, 0.79] & 0.781829 \\
(0.29, 0.3] & 0.697338 & (0.79, 0.8] & 0.788773 \\
(0.3, 0.31] & 0.699306 & (0.8, 0.81] & 0.853704 \\
(0.31, 0.32] & 0.710648 & (0.81, 0.82] & 0.796065 \\
(0.32, 0.33] & 0.725231 & (0.82, 0.83] & 0.742708 \\
(0.33, 0.34] & 0.739005 & (0.83, 0.84] & 0.736921 \\
(0.34, 0.35] & 0.743171 & (0.84, 0.85] & 0.737269 \\
(0.35, 0.36] & 0.714699 & (0.85, 0.86] & 0.712847 \\
(0.36, 0.37] & 0.698148 & (0.86, 0.87] & 0.658333 \\
(0.37, 0.38] & 0.689120 & (0.87, 0.88] & 0.618519 \\
(0.38, 0.39] & 0.660648 & (0.88, 0.89] & 0.622222 \\
(0.39, 0.4] & 0.659838 & (0.89, 0.9] & 0.626389 \\
(0.4, 0.41] & 0.663194 & (0.9, 0.91] & 0.620833 \\
(0.41, 0.42] & 0.646644 & (0.91, 0.92] & 0.655556 \\
(0.42, 0.43] & 0.625000 & (0.92, 0.93] & 0.687500 \\
(0.43, 0.44] & 0.625000 & (0.93, 0.94] & 0.686111 \\
(0.44, 0.45] & 0.626389 & (0.94, 0.95] & 0.739120 \\
(0.45, 0.46] & 0.616667 & (0.95, 0.96] & 0.879514 \\
(0.46, 0.47] & 0.625000 & (0.96, 0.97] & 0.872222 \\
(0.47, 0.48] & 0.625000 & (0.97, 0.98] & 0.864352 \\
(0.48, 0.49] & 0.622222 & (0.98, 0.99] & 0.852315 \\
(0.49, 0.5] & 0.620833 & (0.99, 1.0] & 0.858796 \\
\bottomrule
\end{tabular}
\footnotetext{The unbinned phase coverage table is provided online.}
\end{table}

\begin{table}[h]
\caption{TESS and CHEOPS flares.}\label{ExtDat_flares}%
\begin{tabular}{@{}llllll@{}}
\toprule
Mission & $t_{\rm peak}$ [BJD] & $a$ & $\log_{10} E$ [erg] & orb. phase & ~ \\
\midrule
CHEOPS & 2460392.789 & 0.048 & 35.9$_{-0.04}^{+0.03}$ & 0.025884 &   \\
CHEOPS & 2460400.003 & 0.012 & 34.9$_{-0.04}^{+0.04}$ & 0.062428 &   \\
CHEOPS & 2460455.637 & 0.003 & 34.1$_{-0.12}^{+0.09}$ & 0.056482 &   \\
CHEOPS & 2460413.181 & 0.002 & 33.3$_{-0.08}^{+0.07}$ & 0.955996 & excluded$^1$ \\
TESS & 2458608.290 & 0.005 & 34.6$_{-0.05}^{+0.04}$ & 0.612894 &   \\
TESS & 2458609.676 & 0.011 & 34.7$_{-0.04}^{+0.04}$ & 0.812154 &   \\
TESS & 2459334.890 & 0.010 & 35.2$_{-0.08}^{+0.07}$ & 0.017420 &   \\
TESS & 2459334.981 & 0.024 & 35.5$_{-0.05}^{+0.05}$ & 0.030477 &   \\
TESS & 2459335.173 & 0.006 & 34.9$_{-0.14}^{+0.10}$ & 0.058039 &   \\
TESS & 2459342.014 & 0.008 & 34.6$_{-0.05}^{+0.04}$ & 0.041006 &   \\
TESS & 2460042.757 & 0.005 & 34.3$_{-0.07}^{+0.06}$ & 0.730162 &   \\
TESS & 2460044.282 & 0.005 & 34.5$_{-0.05}^{+0.05}$ & 0.949207 &   \\
TESS & 2460045.424 & 0.003 & 34.0$_{-0.11}^{+0.09}$ & 0.113382 &   \\
TESS & 2460052.991 & 0.004 & 34.3$_{-0.06}^{+0.05}$ & 0.200551 &   \\
TESS & 2460059.608 & 0.049 & 35.5$_{-0.04}^{+0.04}$ & 0.151479 &   \\
TESS & 2460065.829 & 0.004 & 34.2$_{-0.07}^{+0.06}$ & 0.045315 &   \\
\botrule
\end{tabular}

\footnotetext[1]{Flare was excluded from the statistical analysis because it was below the detection threshold of TESS.}
\end{table}

\begin{figure}[h!]
\centering
\includegraphics[width=1.1\textwidth]{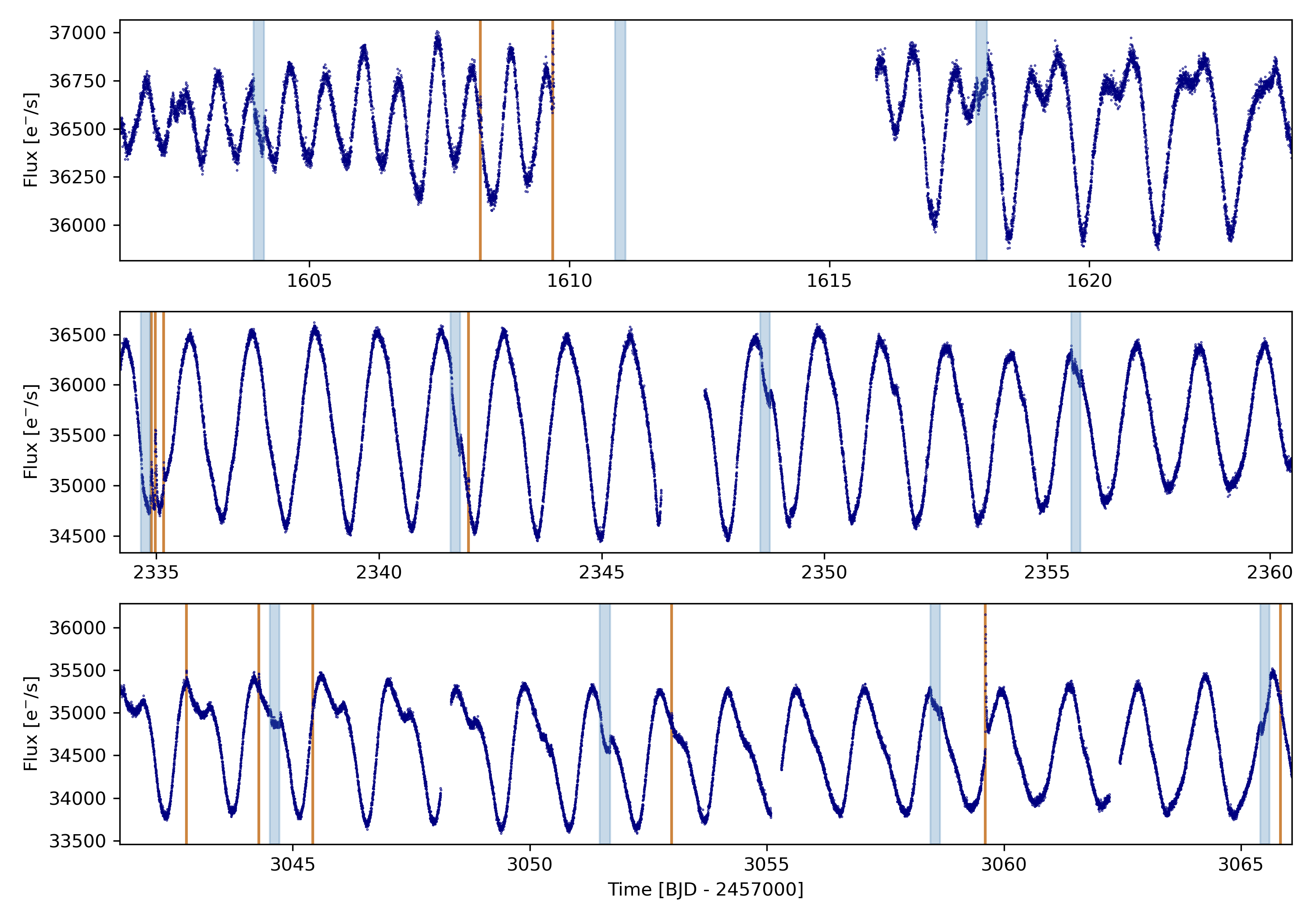}
\caption{TESS light curves of \hip\hspace{-0.1cm}. Blue scatter shows the 2-min TESS \texttt{PDC\_SAP} flux. Light blue shades show the \texttt{batman}~\citep{kreidberg2015batman} transit models of \hip\,b. Orange vertical lines mark the detected flares.}\label{ExtDat_fig:tess}
\end{figure}

\begin{figure}[h!]
\centering
\includegraphics[width=0.9\textwidth]{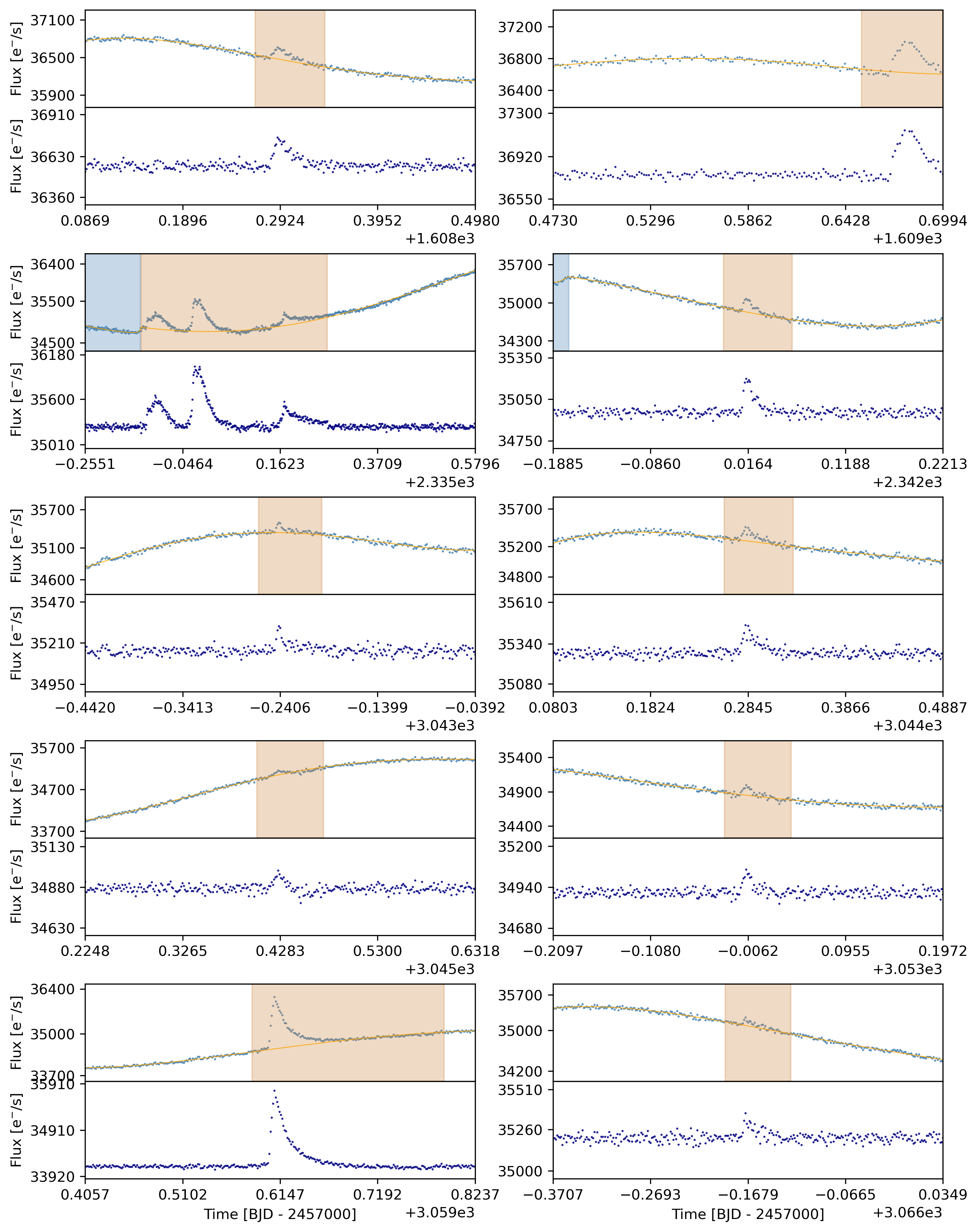}
\caption{Flares in TESS light curves of \hip. Yellow lines show the polynomial fit including the \texttt{batman}~\citep{kreidberg2015batman} transit model of \hip\,b, highlighted with a blue shade. Each bottom panel shows the residual light curve with the polynomial model and transit removed. The flare template was fit within the orange highlighted region.}\label{ExtDat_fig:tessflares}
\end{figure}

\begin{figure}[h!]
 \centering
\includegraphics[width=\textwidth]{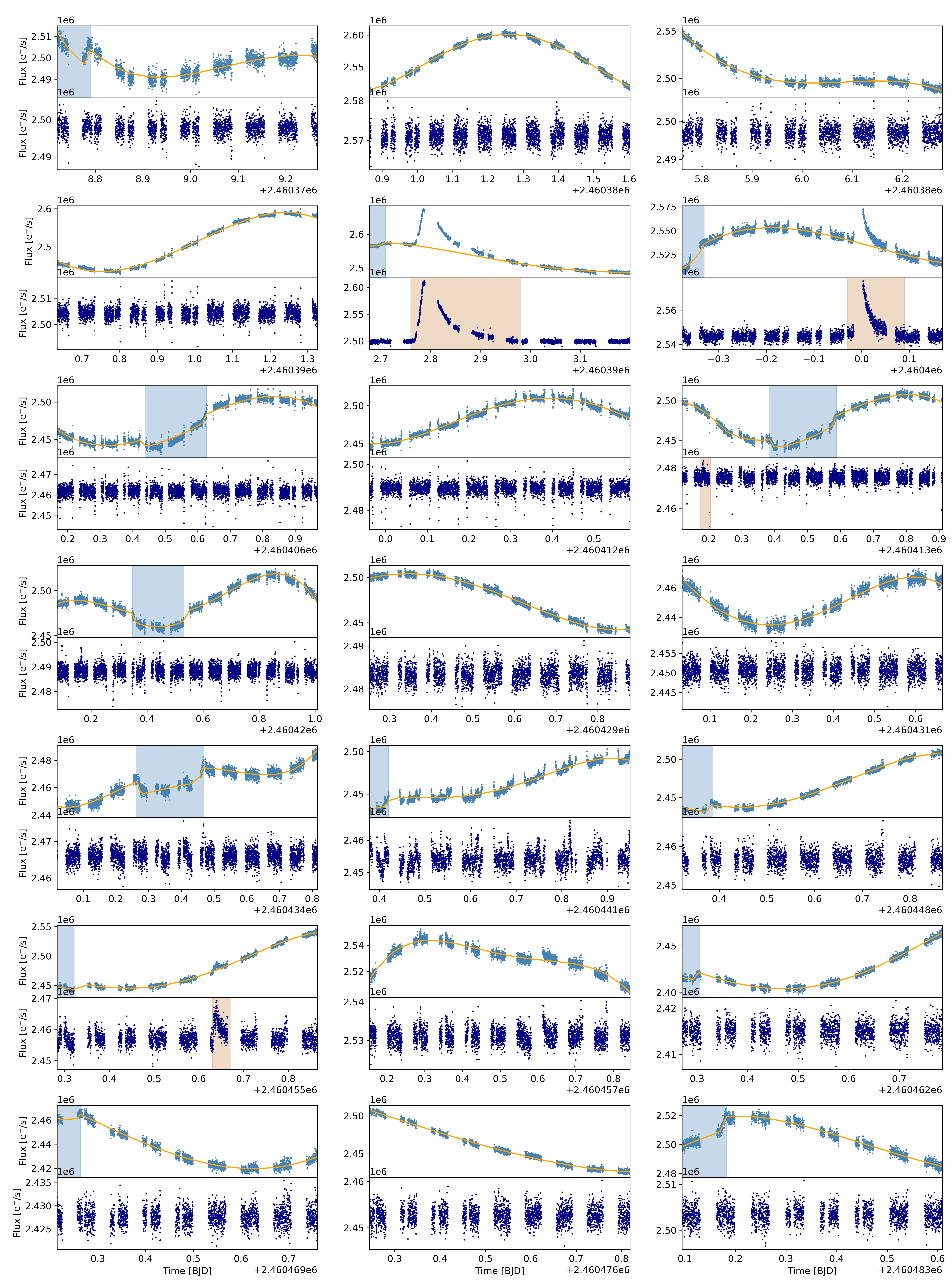}
\caption{CHEOPS light curves of \hip\hspace{-01.cm}. Top panels show the PIPE reduced 10-s cadence imagette light curves as blue scatter.  Yellow lines show the polynomial fit including the \texttt{batman}~\citep{kreidberg2015batman} transit model of \hip\,b, highlighted with a blue shade. Each bottom panel highlight flares within the orange shaded regions, which are reintroduced into the residual light curve (blue scatter).}\label{ExtDat_fig:cheops}
\end{figure}

\begin{figure}[h!]
\centering
\includegraphics[width=0.65\textwidth]{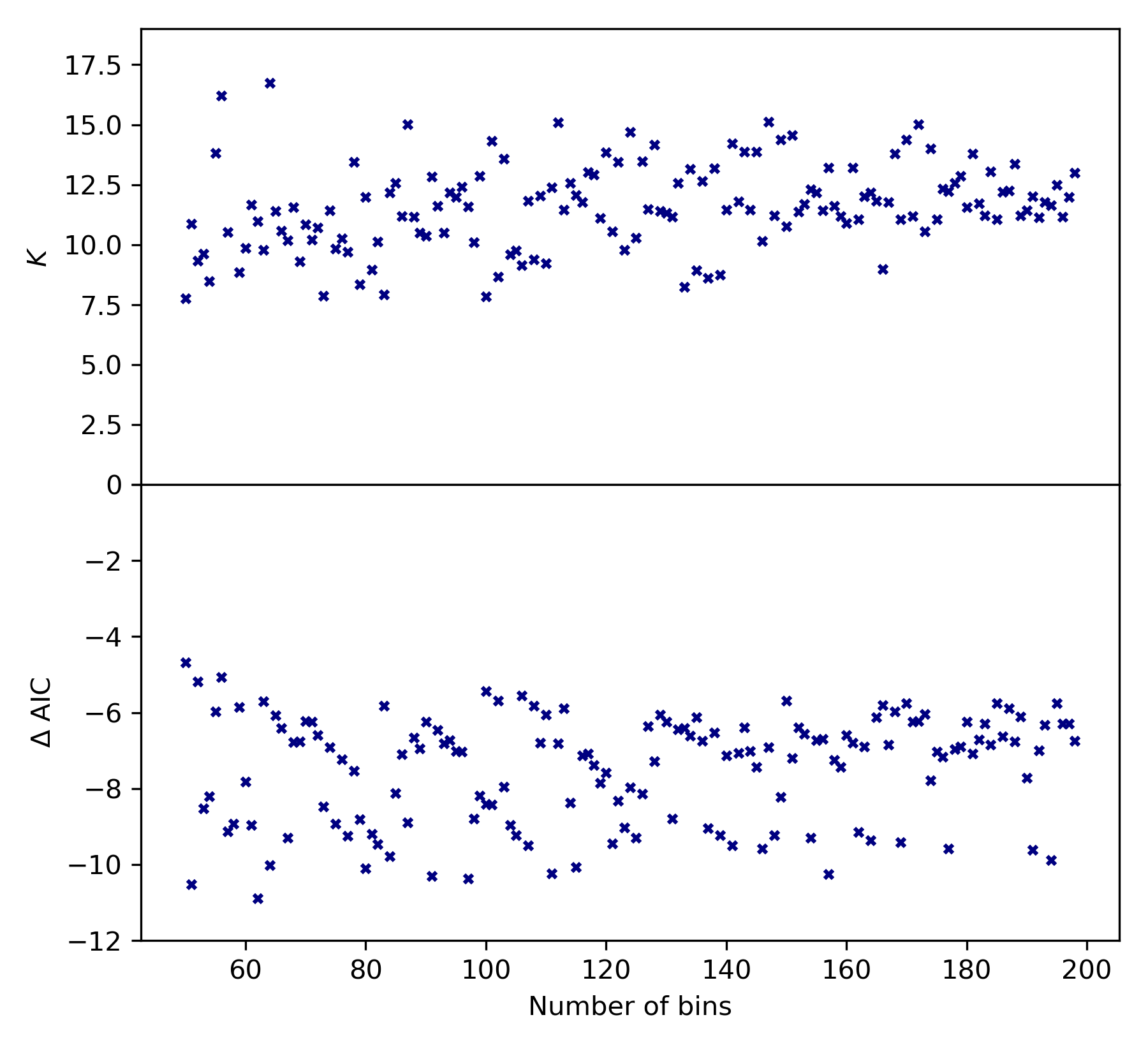}
\caption{Bayes Factor and Akaike Information Criterion as a function of number of orbital phase bins. The values for both statistics converge above 75 bins, indicating that the phases are sufficiently resolved.}\label{ExtDat_fig:kaic}
\end{figure}

\begin{figure}[h!]
\centering
\includegraphics[width=0.65\textwidth]{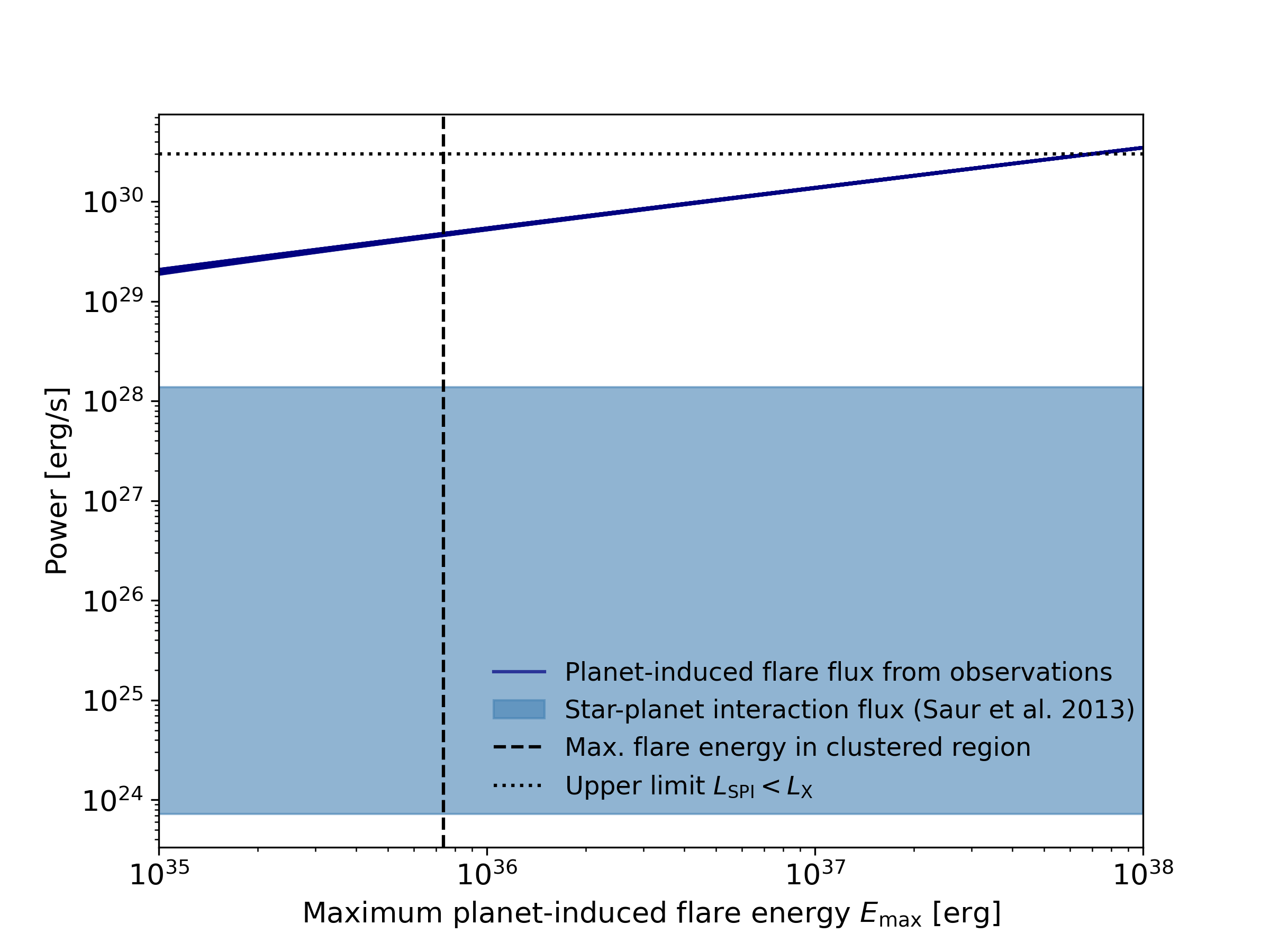}
\caption{Estimate of planet-induced power (overlapping blue solid lines) as a function of highest possible induced flare energy. Lower lines indicate a smaller lower limit for the energy of planet-induced flares in the range $10^{32}-10^{33}$\,erg. Below this limit, flares cannot be distinguished from X-ray quiescent emission~\citep{maggio2024xuv} (dotted line). The measured power is incompatible with the power expected from the Alfv\'en wing star-planet interaction mechanism~\citep{saur2013magnetic} (blue shade), and must therefore tap into a different energy source, e.g. the energy stored in pre-flare coronal loops.}\label{ExtDat_fig:spiflux}
\end{figure}

\begin{figure}[h!]
\centering
\includegraphics[width=0.65\textwidth]{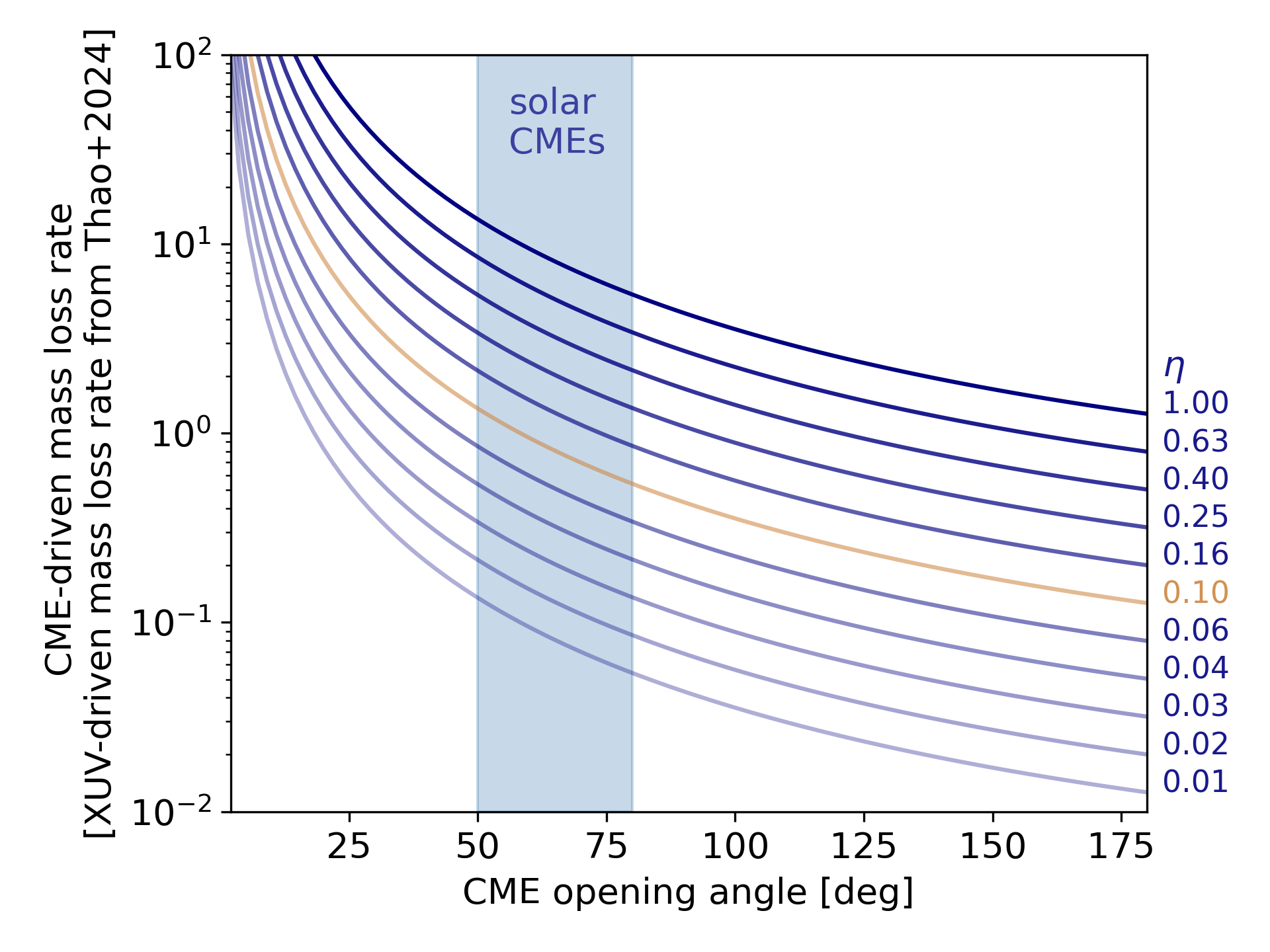}
\caption{Estimate of mass loss rate from \hip\,b due to CMEs as a function of opening angle and efficiency $\eta$, assuming all CMEs intercept the planet's location. The shaded region marks the range of typical opening angles of solar CMEs~\citep{jang2016comparison}. The orange line highlights a typically assumed efficiency factor $
\eta=0.1$.}\label{ExtDat_fig:massloss}
\end{figure}

\begin{figure}[h!]
\centering
\includegraphics[width=0.65\textwidth]{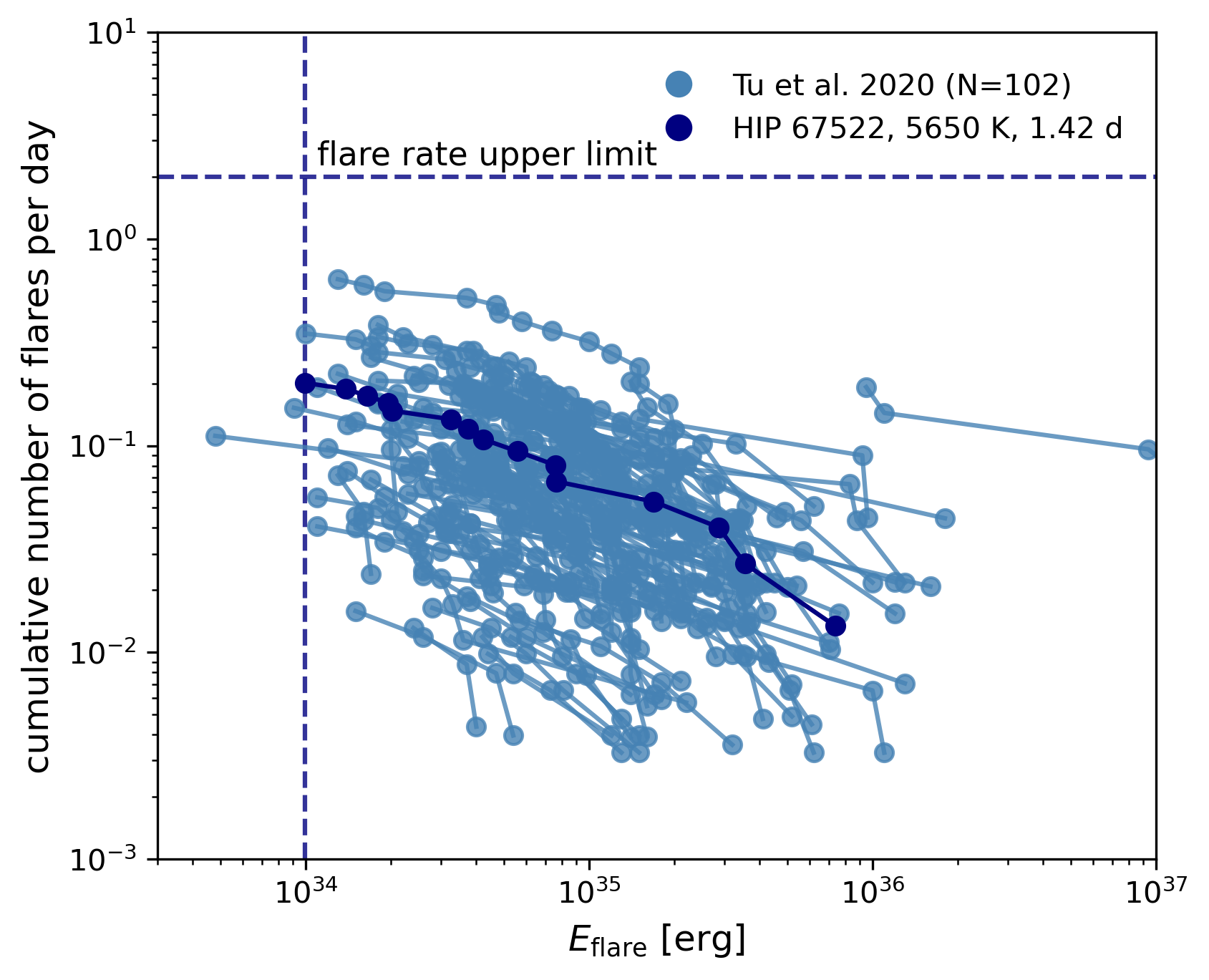}
\caption{Flare frequency distributions of G dwarf stars with rotation periods $0.27-5.41\,$d, and effective temperatures $5113- 5916\,$K with at least three flares observed with TESS~\citep{tu2020superflares}, excluding known eclipsing binaries from \cite{ziegler2021soar}. We set the upper limit on the flare rates $\lambda_0$ and $\lambda_1$ conservatively to $2\,$\perday~(horizontal dashed line) above the minimum flare energy in the sample~(vertical dashed line). Note that the \cite{tu2020superflares} sample still likely contains binary stars, such that the true flare rate of some stars is lower than shown.}\label{ExtDat_fig:tu}
\end{figure}




\end{appendices}

\newpage
\bibliography{sn-bibliography}

\end{document}